\documentclass[lettersize,journal]{IEEEtran}
\usepackage{amsmath,amsfonts}
\usepackage{algorithmic}
\usepackage{algorithm}
\usepackage{array}
\usepackage[caption=false,font=normalsize,labelfont=sf,textfont=sf]{subfig}
\usepackage{textcomp}
\usepackage{stfloats}
\usepackage{url}
\usepackage{verbatim}
\usepackage{graphicx}
\usepackage{cite}
\usepackage{xcolor}
\usepackage[binary-units]{siunitx}
\usepackage{orcidlink}
\begin{document}

\newcommand{\chipnames}{SpiDR}
\newcommand{\chipname}{SpiDR }

% \title{A Scalable and Reconfigurable Digital Compute-in Memory Spiking Neural Network Accelerator Core for Event-based Perception}
\title{\chipnames: A Reconfigurable Digital Compute-in-Memory Spiking Neural Network Accelerator for Event-based Perception}

%Possible acronym: ERISED, DESIRE, SCARE, NIMBLE, SpiRE, ReSpiCE

\author{Deepika Sharma \orcidlink{0000-0002-8244-7919},~\IEEEmembership{Graduate Student Member,~IEEE,} Shubham Negi \orcidlink{0000-0001-7413-9636},~\IEEEmembership{Graduate Student Member,~IEEE,} \\ Trishit Dutta \orcidlink{0009-0003-7659-9531},~\IEEEmembership{Member,~IEEE,} Amogh Agrawal \orcidlink{0000-0001-9999-9085},~\IEEEmembership{Member,~IEEE,} and Kaushik Roy \orcidlink{0000-0002-0735-9695},~\IEEEmembership{Fellow,~IEEE,}
        % <-this % stops a space
% \thanks{This paper was produced by the IEEE Publication Technology Group. They are in Piscataway, NJ.}% <-this % stops a space
% \thanks{Manuscript received April 19, 2021; revised August 16, 2021.}
% \thanks{Manuscript received ....}
}

% \author{IEEE Publication Technology,~\IEEEmembership{Staff,~IEEE,}
%         % <-this % stops a space
% \thanks{This paper was produced by the IEEE Publication Technology Group. They are in Piscataway, NJ.}% <-this % stops a space
% \thanks{Manuscript received April 19, 2021; revised August 16, 2021.}}

% The paper headers
\markboth{Journal of \LaTeX\ Class Files,~Vol.~xx, No.~xx, Month~Year}%
{Shell \MakeLowercase{\textit{et al.}}: A Sample Article Using IEEEtran.cls goofor IEEE Journals}

\IEEEpubid{0000--0000/00\$00.00~\copyright~2024 IEEE}
% Remember, if you use this you must call \IEEEpubidadjcol in the second
% column for its text to clear the IEEEpubid mark.

% {An Input Sparsity-aware Scalable and Reconfigurable Digital Compute-in Memory Spiking Neural Network Accelerator Core for Event-based Perception}

\maketitle

\begin{abstract}
Spiking Neural Networks (SNNs), with their inherent recurrence, offer an efficient method for processing the asynchronous temporal data generated by Dynamic Vision Sensors (DVS), making them well-suited for event-based vision applications. However, existing SNN accelerators suffer from limitations in adaptability to diverse neuron models, bit precisions and network sizes, inefficient membrane potential (Vmem) handling, and limited sparse optimizations. In response to these challenges, we propose a scalable and reconfigurable digital compute-in-memory (CIM) SNN accelerator \chipname with a set of key features: 1) It uses in-memory computations and reconfigurable operating modes to minimize data movement associated with weight and Vmem data structures while efficiently adapting to different workloads. 2) It supports multiple weight/Vmem bit precision values, enabling a trade-off between accuracy and energy efficiency and enhancing adaptability to diverse application demands. 3) A zero-skipping mechanism for sparse inputs significantly reduces energy usage by leveraging the inherent sparsity of spikes without introducing high overheads for low sparsity. 4) Finally, the asynchronous handshaking mechanism maintains the computational efficiency of the pipeline for variable execution times of different computation units. We fabricated \chipname in 65 nm Taiwan Semiconductor Manufacturing Company (TSMC) low-power (LP) technology. It demonstrates competitive performance (scaled to the same technology node) to other digital SNN accelerators proposed in the recent literature and supports advanced reconfigurability. It achieves up to 5 TOPS/W energy efficiency at 95\% input sparsity with 4-bit weights and 7-bit Vmem precision.
\end{abstract}


\section{Introduction}\label{sec:intro}
% Things that we need to talk about here:
    % background, motivation, related works
    % SNNs
    % digital in-memory computing

\begin{figure}[!t]
\centering
\includegraphics[width=\linewidth]{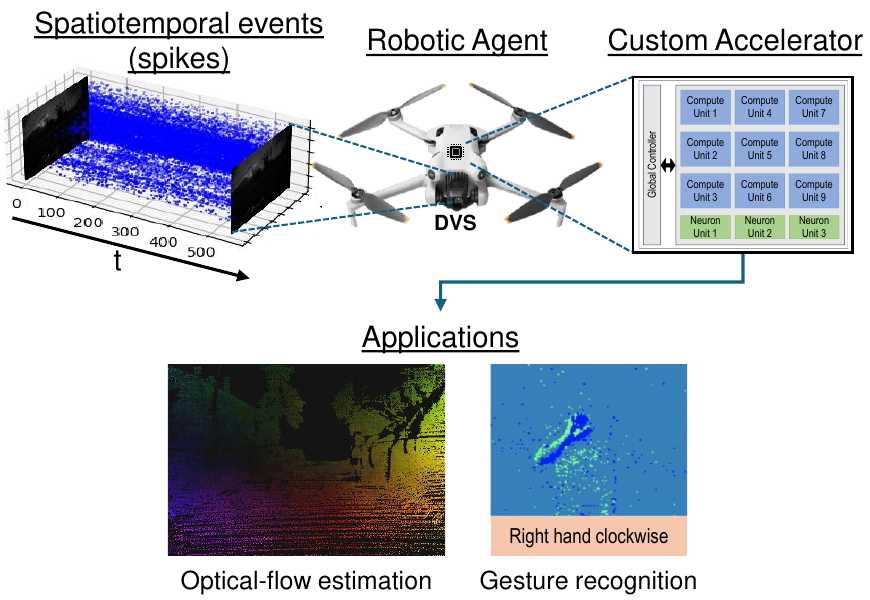}
\caption{A conceptual diagram showcasing the use of DVS camera and a specialized accelerator for real-time motion analysis tasks (such as optical-flow estimation) in a robotic agent.}
\label{fig:event_pipeline}
\end{figure}

% About DVS and their applications
In the evolving landscape of machine vision, dynamic vision sensors (DVS)~\cite{lichtsteiner2008128, brandli2014240} have emerged as a unique visual sensing paradigm. Their event-based operation, where pixels asynchronously output data only upon detecting changes in intensity, offers inherent advantages over traditional frame-based cameras. This approach reduces data redundancy and enables high temporal resolution, which makes DVS particularly suitable for dynamic and fast-changing environments~\cite{gallego2020event}. Among the numerous applications benefiting from DVS technology, object detection, tracking, optical flow estimation, etc. are critical for real-time motion analysis, robotics, and autonomous navigation systems~\cite{gallego2020event}. Fig.~\ref{fig:event_pipeline} provides an overview of a DVS-based system for real-time motion analysis in a robotic agent.

% Processing DVS data  -> ANNs (not efficient and lose temporal information) -> SNNs come to the rescue -> traditional general-purpose computing architectures don't support SNNs very efficiently -> Therefore we need custom/specialized accelerators
Nevertheless, processing Dynamic Vision Sensor (DVS) data introduces unique challenges and opportunities, primarily due to its event-driven and sparse nature. Traditional Analog Neural Networks (ANNs)\footnote{ANN refers to standard neural networks that operate on multi-bit input and activation values (unlike spike-based digital inputs), including CNNs, MLPs, RNNs, LSTMs, etc.}, while powerful for a wide range of applications, face difficulties fully utilizing the asynchronous event streams and precise temporal information provided by DVS data. In contrast, Spiking Neural Networks (SNNs), inspired by biological neural dynamics, inherently align with the sparse, event-driven nature of DVS data and offer a more efficient processing paradigm~\cite{maass1997networks, kosta2023adaptive}.

\IEEEpubidadjcol

However, traditional general-purpose computing architectures, such as CPUs and GPUs, struggle to accommodate the sparse and asynchronous nature of SNN computations~\cite{parker2022benchmarking}. Designed around synchronous data flows, they incur overheads when converting event-based data into frame-like representations. In addition, their regular memory access patterns do not align with the unpredictable dynamic computations of SNNs, leading to an underutilization of processing units during sparse computations. Therefore, there is increased interest in developing specialized accelerators tailored to the unique demands of event-based processing to maximize performance and efficiency~\cite{basu2022spiking, kim2023c, zhang202322, frenkel2022reckon,  kim2022neuro, chang202373, liu2023aa}.

\begin{figure}[!t]
\centering
\includegraphics[width=\linewidth]{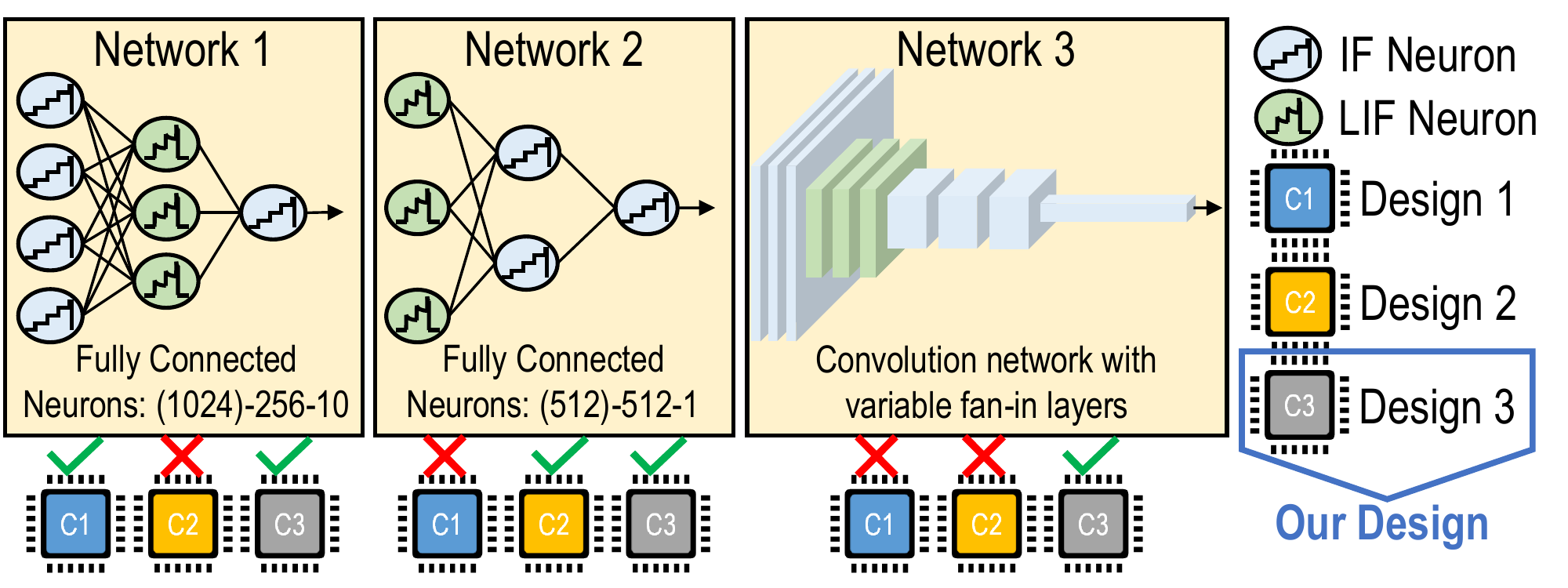}
\caption{Lack of reconfigurability in existing SNN architectures.}
\label{fig:limit1}
\end{figure}

% Limitations of previous recent chips/designs -> What do we offer to take care of those limitations (simultaneously introduce our contributions)
Despite these efforts, current SNN accelerators face several limitations in terms of scalability and flexibility that hinder their widespread adoption. A key limitation among these is the lack of versatile support for diverse network architectures~\cite{zhang202322, frenkel2022reckon, stuijt2021mubrain, park20197}, variable bit precision~\cite{zhang202322, frenkel2022reckon, stuijt2021mubrain} and different neuron models~\cite{kim2023c, zhang202322, frenkel2022reckon, kim2022neuro, park20197} (Fig.~\ref{fig:limit1}). Limitations in architectural support restrict the types of networks that can be efficiently implemented on the hardware. Furthermore, bit precision constraints hinder the exploration of accuracy and energy consumption trade-offs, and the inability to accommodate different neuron models limits the computational capabilities of SNNs on these accelerators. Additionally, many existing designs do not discuss the storage and data movement overhead of the membrane potential (Vmem) data structure. As a result, even when low precision is used for weights, high precision is used for Vmems~\cite{kim2023c, zhang202322, liu2023aa}, which can add high processing and memory overhead, particularly for layers with large output dimensions.

\begin{figure}[!t]
\centering
\includegraphics[width=\linewidth]{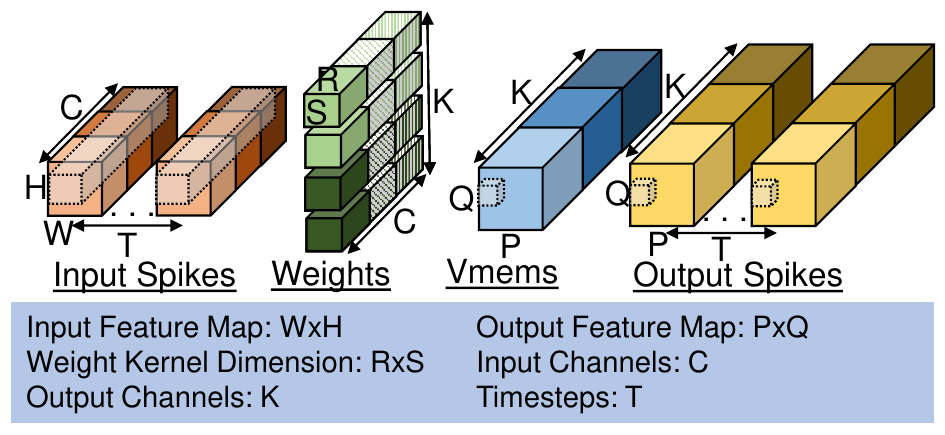}
\caption{An example spiking convolution layer.}
\label{fig:sp_conv_layer}
\end{figure}

\begin{figure}[!t]
\centering
\includegraphics[width=\linewidth]{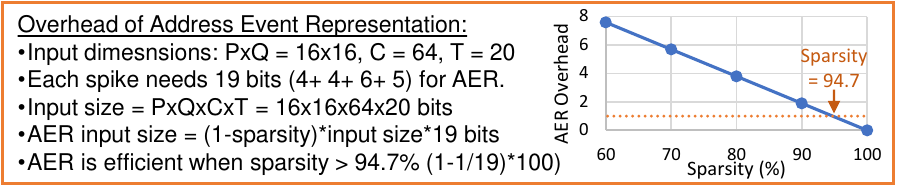}
\caption{Overhead of using AER for input spikes for varying input sparsity.}
\label{fig:AER_overhead}
\end{figure}

\begin{figure}[!t]
\centering
\subfloat[Gesture recognition]{\includegraphics[width=1.7in]{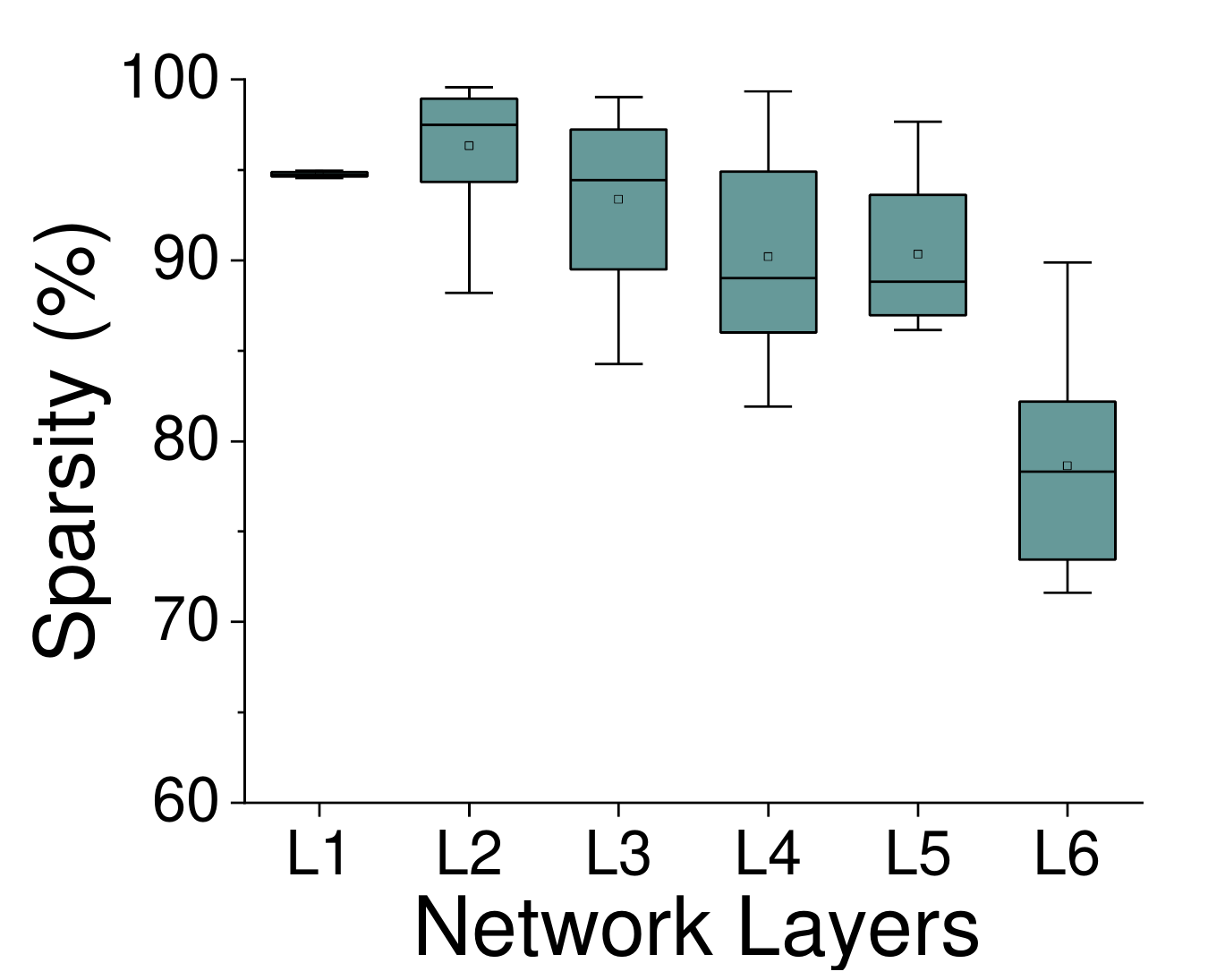}%
\label{fig_first_case}}
\hfil
\subfloat[Optical flow estimation]{\includegraphics[width=1.7in]{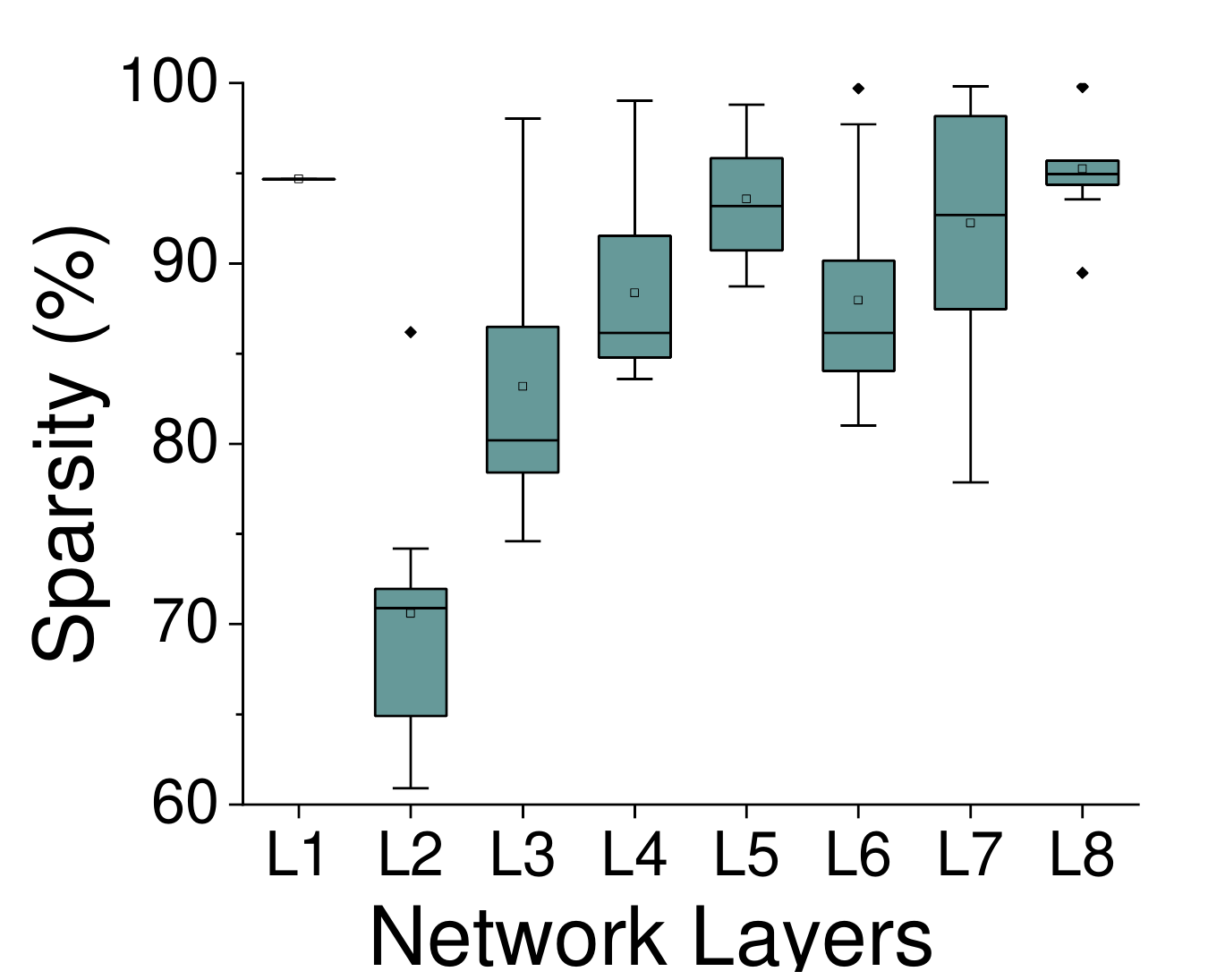}%
\label{fig_second_case}}
\caption{Variation in input sparsity across different networks and layers of the same network.}
\label{fig:sparsity}
\end{figure}

Finally, to achieve high efficiency, most SNN accelerators rely on sparse optimizations, such as address event representation (AER) for input spikes~\cite{zhang202322, frenkel2022reckon}. However, AER is primarily beneficial for data structures with a high sparsity or small size (fewer data points). This leads to inconsistencies since the sparsity levels of input spikes in SNNs can vary significantly across different network architectures or even within layers of the same network. Fig.~\ref{fig:sp_conv_layer} illustrates the various data structures in a general spiking convolution layer. Fig.~\ref{fig:AER_overhead} then highlights the overhead of using AER at different input sparsity levels for such a layer, demonstrating that AER is effective only when sparsity exceeds 94.7\% in this example. Fig.~\ref{fig:sparsity} further reinforces our point by showcasing the variation in sparsity across layers in two SNNs designed for gesture recognition and optical flow estimation tasks. The details of these networks are presented in Section~\ref{sec:experiments}. Notably, the sparsity of inputs for the second layer of the optical flow estimation network can be as low as 60\% and never goes beyond 75\% whereas, the sparsity of the next layer inputs ranges from 75\% to 99\%. This analysis underscores the need for SNN accelerators that can efficiently handle varying sparsity levels rather than relying solely on optimizations that are effective in specific cases but increase the overheads in other cases.
% Such a large variation in sparsity shows that any sparsity optimization targeted at a narrow sparsity range will not be beneficial for a wide range of other sparsity levels.

\begin{figure}[!t]
\centering
\includegraphics[width=0.9\linewidth]{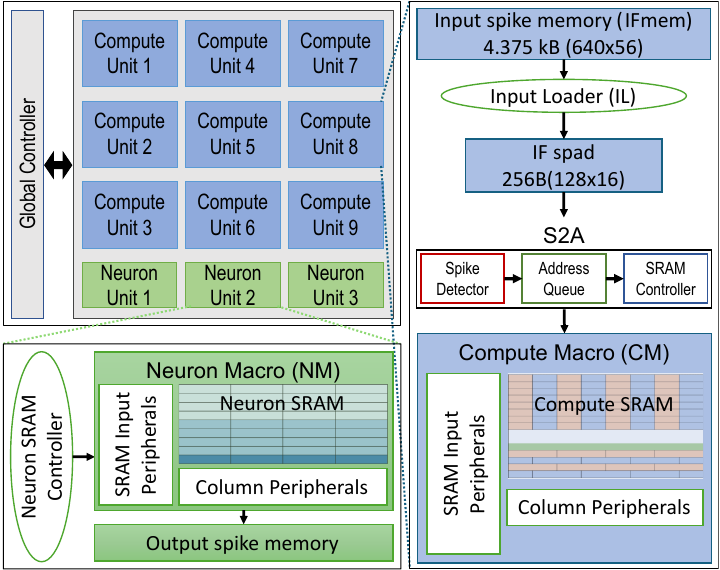}
\caption{Proposed SNN core architecture}
\label{fig:core_arch}
\end{figure}

This work introduces a scalable and reconfigurable SNN inference core \chipnames, that addresses the limitations discussed above through several key strategies outlined below:
\begin{itemize}
    \item It uses in-memory computation, reconfigurable operating modes, and timestep pipelining to efficiently support various workload sizes while minimizing weight and Vmem data movement.
    \item It seamlessly incorporates three weight/Vmem bit precisions (4/7, 6/11, 8/15) into the macro design, enhancing parallelism with full resource utilization, allowing a balanced trade-off between computational accuracy and energy consumption.
    \item It employs a zero-skipping strategy to leverage a wide range of input sparsity, reducing energy and latency with minimal overhead.
    \item It uses an asynchronous handshaking mechanism to maintain the efficiency of the computation pipeline.
\end{itemize}

\section{Proposed SNN Core Architecture}\label{sec:architecture}

Fig.~\ref{fig:core_arch} shows the proposed SNN core with 9 compute units (CU) and 3 neuron units (NU). The compute units accumulate weights into partial Vmems based on input spike values. Each compute unit consists of an input spike memory (IFmem), input loader (IL), input scratchpad buffer (IFspad), spike-to-address converter (S2A), and a CIM compute macro (CM). The IFmem stores the input spike values in raw/uncompressed form. The input loader transfers data from IFmem to IFspad, aligning for convolution (Conv) or fully connected (FC) layer operations. The spike-to-address converter converts IFspad spikes into weight and Vmem row addresses for in-memory weight to Vmem accumulation in the compute macro. Neuron units receive partial Vmems from compute units and perform the required neuron operations (e.g. accumulation and threshold comparison for Integrate and Fire (IF) neuron) using the CIM neuron macro (NM) coordinated by a neuron SRAM controller. The following paragraphs in this section describe each of these components in detail.

\subsection{Macro design}\label{sec:macro}
\begin{figure}[!t]
\centering
\includegraphics[width=\linewidth]{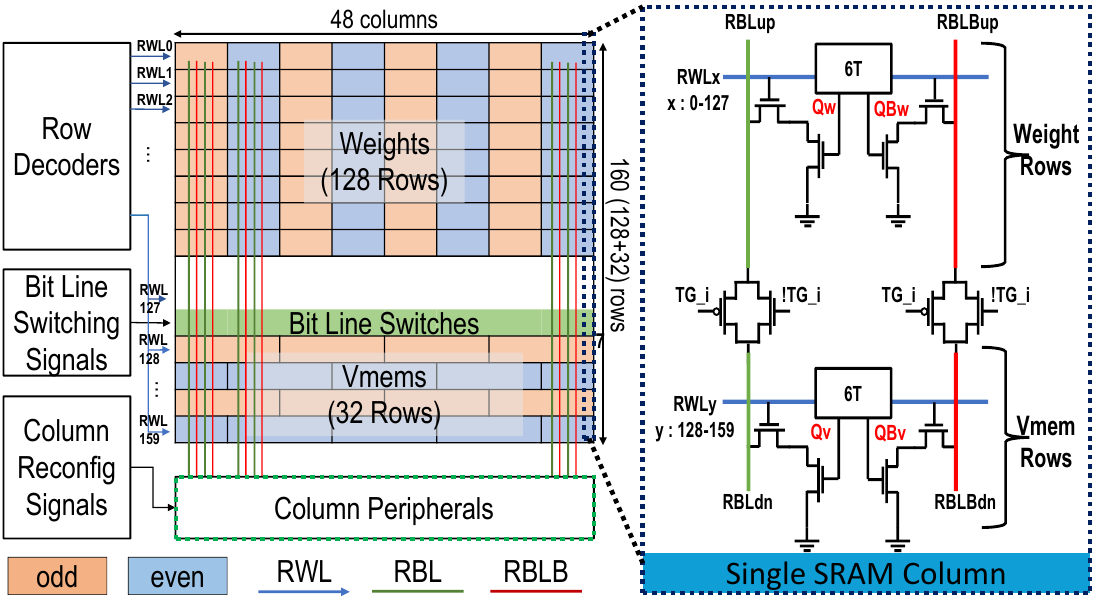}
\caption{Compute macro architecture.}
\label{fig:comp_macro}
\end{figure}

% \begin{figure}[!t]
% \centering
% \includegraphics[width=1in]{figures/column_peripherals.pdf}
% \caption{Column peripherals.}
% \label{fig:col_peri}
% \end{figure}

% \begin{figure}[!t]
% \centering
% \includegraphics[width=3in]{figures/bit_prec.pdf}
% \caption{Variable bit precision support.}
% \label{fig:bit_prec}
% \end{figure}

Fig.~\ref{fig:comp_macro} details the compute macro, an enhanced version of the design presented in IMPULSE~\cite{agrawal2021impulse}. The core component of the macro is a \(160\times48\) 10T SRAM array, where the top 128 rows store synaptic weights, and the remaining 32 rows are used for storing partial membrane potentials (Vmems). Similar to IMPULSE, this design integrates the weight and Vmem arrays within the same SRAM structure. This co-location enables the accumulation of weights into partial Vmems directly within the memory array using some peripheral circuitry. This in-memory approach significantly reduces the energy-intensive data movement typically required between separate memory and compute units, thereby enhancing the overall efficiency of the SNN accelerator.

As multiple weights accumulate into Vmems, the precision required for Vmems is usually greater than the precision required for weights. In this work, we assume the Vmem precision (\(B_{Vmem}\)) to be roughly equal to two times the weight precision (\(B_{weight}\)), that is, \(B_{Vmem} = 2*B_{weight}-1\). Consequently, we need two memory rows to store the Vmems corresponding to one weight row in a staggered layout. IMPULSE~\cite{agrawal2021impulse} uses an additional read word line (RWL) for each weight row to support these staggered computations. A key enhancement in our macro design is using transmission gates as read bit line (RBL) switches to connect the bit lines between weight and Vmem arrays. This modification eliminates the need for the additional RWL per weight memory row.

\begin{figure}[!t]
\centering
\subfloat[]{\includegraphics[width=0.67\linewidth]{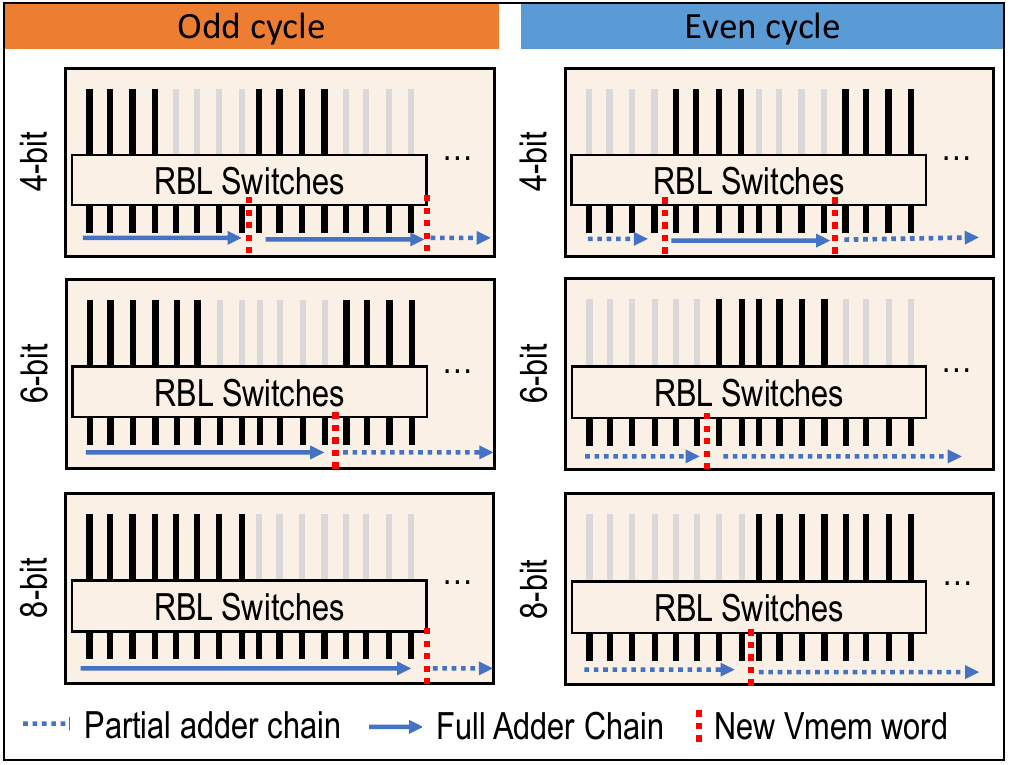}%
\label{fig:bit_prec}}
\hfil
\subfloat[]{\includegraphics[width=0.3\linewidth]{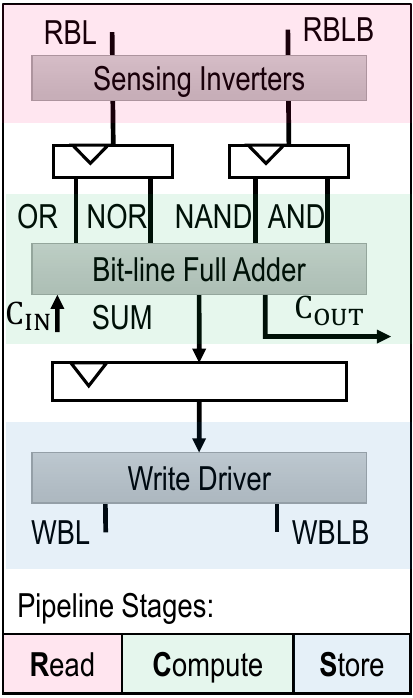}%
\label{fig:col_peri}}
\caption{(a) Varibale bit percision support. (b) Column peripherals.}
\label{fig:comp_macro2}
\end{figure}

Moreover, this design also enables reconfigurable bit precision support. The previous macro~\cite{agrawal2021impulse} was limited to a single weight/Vmem bit precision configuration (6/11-bit) determined at design time by its dual RWL connectivity. Our compute macro offers greater flexibility by supporting multiple weight and Vmem bit precisions by changing the RBL switch connectivity and column peripheral configurations. Supported weight/Vmem bit precision configurations include 4/7-bit, 6/11-bit, and 8/15-bit, enabling the accelerator to adapt to the accuracy and efficiency requirements of different SNN workloads. The bit precision required for a particular workload can be selected as a configuration parameter before starting execution, avoiding any reconfigurability overhead. Fig.~\ref{fig:bit_prec} shows the RBL switch and column adder chain configurations for different bit precisions and even/odd accumulation cycles. Bit lines are numbered in ascending order (0 to 47) from left to right. For a 4-bit odd cycle operation, as an example, odd weights are added to an odd Vmem row, activating RBL switches at lines (0-3), (8-11), ... , and  (40-43).

The weight to Vmem accumulation operation is performed with the help of the column peripheral circuitry illustrated in Fig.~\ref{fig:col_peri}. Adding one row of weights to Vmems consists of three pipelined stages:
\begin{enumerate}
    \item \textbf{Read (R):} One weight RWL and one Vmem RWL are activated. This generates \(NOR\) and \(AND\) outputs at RBL and RBL bar (RBLB), that are sensed using inverters and latched. Latching these outputs allows them to be used reliably in the next compute stage.
    \item \textbf{Compute (C):} The latched outputs, along with the carry-in \((C_{IN})\) bit from the adjacent column, are used to generate the \(SUM\) and carry-out \((C_{OUT})\) bits. A simple adder circuit performs this computation.
    \item \textbf{Store (S):} The \(SUM\) bit is written to the active Vmem row.
\end{enumerate}

% Neuron macros share a similar design, with 72 rows (32 for partial Vmems, 32 for full Vmems, and the rest for the neuron threshold and leak values). The neuron macros support integrate and fire (IF) and leaky integrate and fire (LIF) neuron models with soft and hard reset options. The hard reset option resets a Vmem value to \(0\) after an output spike, whereas the soft reset option resets it to a value equivalent to the difference between Vmem and the threshold voltage \((Vmem - V_{th})\). The \textit{Store} stage of the computation pipeline is augmented to perform this conditional write operation. Unlike IMPULSE~\cite{agrawal2021impulse}, which uses a single macro to perform weight to Vmem accumulation, and the neuron operations, we separate these operations between the compute and neuron macros, respectively. This separation allows us to enable reconfigurable support for variable workload sizes as elaborated in Section~\ref{sec:modes}.

The design of neuron macros is similar to compute macros, with a \(72\times48\) memory array. Of these, \(32\) rows store partial Vmems (membrane potentials received from the compute units), another \(32\) store full Vmems, and the remaining rows are reserved for storing neuron model-specific parameters like thresholds and leak values.

The neuron macros support integrate-and-fire (IF) and leaky-integrate-and-fire (LIF) neuron models with soft and hard reset options following a neuron spike. The hard reset option resets the Vmem to zero, while the soft reset option subtracts the threshold voltage from the Vmem, retaining some residual potential. The choice of neuron model and reset method influences the neuron's firing dynamics and can be tailored to specific SNN tasks. To implement these reset options, the \textit{Store} stage of the compute pipeline is augmented with conditional write logic, allowing it to perform the appropriate reset operation only when a neuron fires (i.e. generates an output spike).

Another significant difference from the IMPULSE design~\cite{agrawal2021impulse} is the separation of functionality. While IMPULSE combined weight-to-Vmem accumulation and neuron operations within a single macro, our design explicitly divides these tasks between the compute and neuron macros. This decoupling enhances the accelerator's flexibility, allowing it to handle SNNs with varying layer sizes efficiently. The reconfigurable operating modes enabled by this separation are discussed in greater detail in Section~\ref{sec:modes}.
% Placeholder for input loading.

\subsection{Amortizing switching overhead}\label{sec:switching}

\begin{figure}[!t]
\centering
\includegraphics[width=0.7\linewidth]{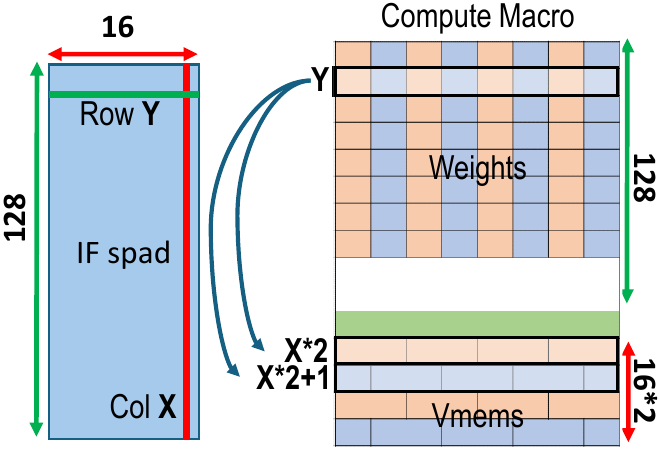}
\caption{Effective operations corresponding to one spike at row Y and column X in IFspad.}
\label{fig:effec_ops}
\end{figure}

\begin{figure}[!t]
\centering
\includegraphics[width=\linewidth]{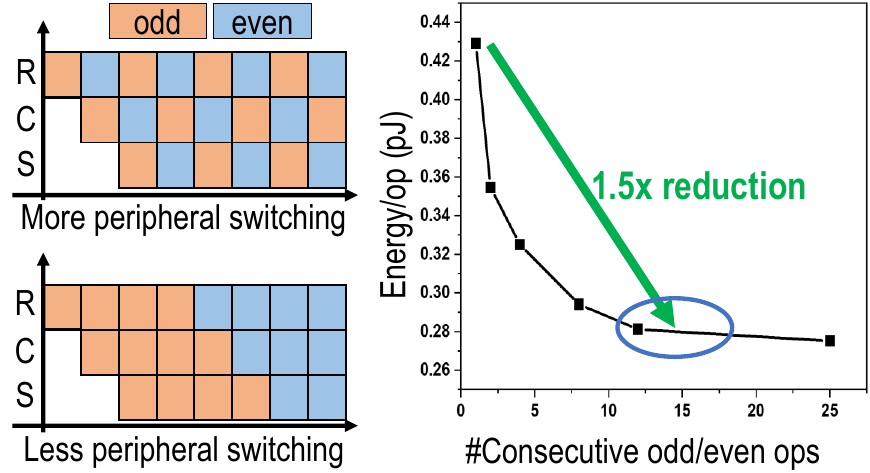}
\caption{Peripheral switching Overhead.}
\label{fig:switching_overhead}
\end{figure}

Fig.~\ref{fig:effec_ops} illustrates the mapping of input spikes in the \(128\times16\) IFspad array with respect to the weights and Vmem rows of the compute macro. Each row (\(Y\)) in the IFspad directly corresponds to a row of the weight memory and each column (\(X\)) of the IFspad is associated with two consecutive rows of the Vmem memory – an even row \(X*2\) and an odd row \(X*2+1\). This mapping facilitates the accumulation of weights into corresponding Vmem locations. The address of each spike (\(Y, X\)) in the IFspad is fed to the address queue. The compute macro then performs two operations for each (\(Y, X\)) pair: 

\begin{itemize}
    \item \textbf{Even Accumulation:} The even-numbered weights from row (\(Y\)) in the weight memory are added to the values stored in the even Vmem row \((X*2)\).
    \item \textbf{Odd Accumulation:} The odd-numbered weights from the same row (\(Y\)) are added to the values in the odd Vmem row \((X*2+1)\).
\end{itemize}

A naive way would be to consecutively perform one even and one odd cycle operation. However, switching between even and odd operations requires changing the bit line switch and peripheral connectivity as shown in Fig.~\ref{fig:bit_prec}. We studied the overhead of this continuous switching on the overall energy consumption of the compute macro (Fig.~\ref{fig:switching_overhead}) and separate the even and odd operations to minimize this overhead. By performing multiple consecutive operations of the same type (even or odd), we can significantly reduce the frequency of these switches, resulting in substantial energy savings. As shown in Fig.~\ref{fig:switching_overhead}, switching peripherals after \(15\) consecutive even or odd operations, as opposed to after every cycle, results in a \(1.5\times\) reduction in energy per operation. This insight drives the design of our spike-to-address converter (S2A), ensuring that spikes are processed with optimal energy savings.

\subsection{Spike-to-address converter (S2A)}\label{sec:s2a}
% The S2A (Fig.~\ref{fig:s2a}) has three components: spike detector, address queue, and SRAM controller. A row from the IFspad, representing weight address \(Y\), is read and processed by a low-complexity spike detector (trailing zero detector) to determine the Vmem address (\(X\)). This address is stored in the address queue as a weight, Vmem address tuple (\(Y, X\)). The SRAM controller reads from the queue, issuing control signals for the compute macro’s IMC operation. To reduce the even-odd switching frequency, we employ an even-odd ping-pong FIFO for the queue. After reading the address tuple from the even FIFO queue and performing the corresponding accumulation, it is written to the odd FIFO queue. The operating mode switches from even to odd once the odd FIFO is full or the even FIFO is empty. The depth of the even and odd FIFOs is set to 16, as significant energy savings are observed only up to this reduced switching frequency (Fig.~\ref{fig:switching_overhead}).

\begin{figure}[!t]
\centering
\includegraphics[width=\linewidth]{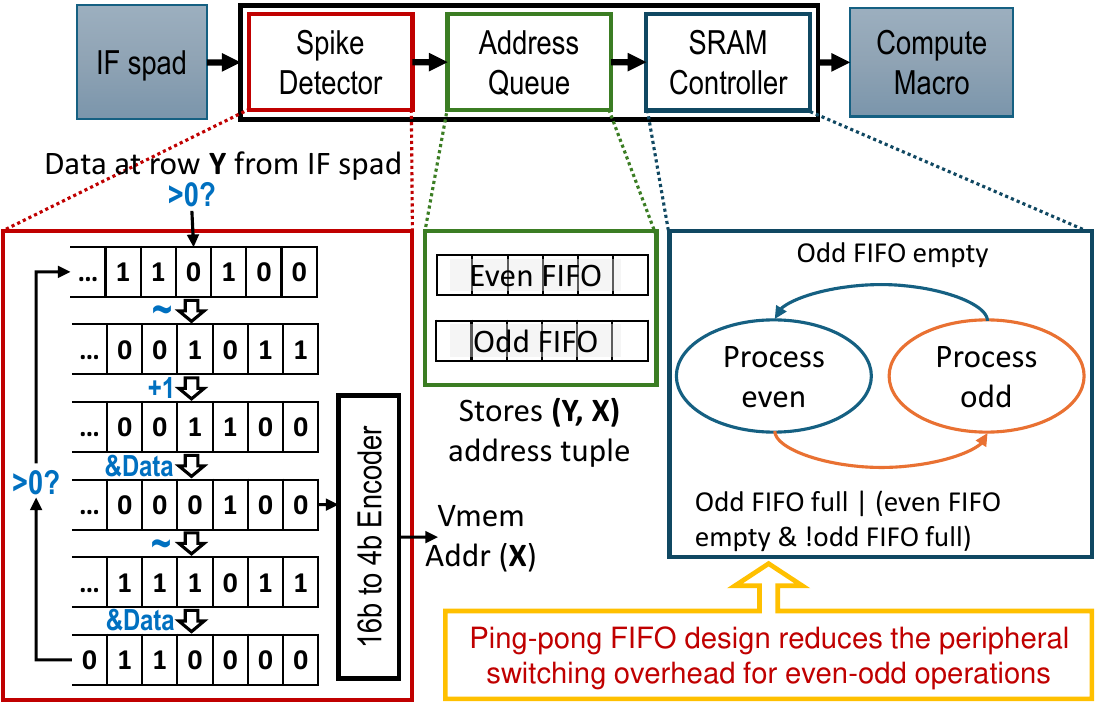}
\caption{Peripheral switching Overhead.}
\label{fig:s2a}
\end{figure}

The S2A (Fig.~\ref{fig:s2a}) consists of three primary components: a spike detector, an address queue, and an SRAM controller. The spike detector, a low-complexity trailing zero detector, reads and processes a row of the IFspad, representing weight address \(Y\), and identifies the column location of the spikes, representing Vmem address (\(X\)). This address is stored in the address queue as a weight, Vmem address tuple (\(Y, X\)). The SRAM controller reads the address tuple from the address queue and issues control signals to the compute macro to perform the accumulation operation. 

The address queue is designed to minimize the even-odd operation switching as explained in Section~\ref{sec:switching}. We employ an even-odd ping-pong FIFO mechanism for the address queue. This means that after reading and processing an address tuple from the even FIFO, the same tuple is written to the odd FIFO. This mechanism strategically batches consecutive even or odd operations, as shown in Fig.~\ref{fig:switching_overhead}. The SRAM controller oversees this ping-pong behavior, switching between even and odd operations only when the current FIFO is empty or the other FIFO is full. The depth of both even and odd FIFOs is set to 16, as a further increase in depth does not provide significant additional energy reduction, as evidenced by Fig.~\ref{fig:switching_overhead}.

\subsection{Input loader}\label{sec:input_loader}
We use a dual-port SRAM to implement the IFspad, allowing simultaneous read and write operations. The spike detector in the spike-to-address converter (S2A) is connected to the read port, while the input loader (IL) writes new data into the IFspad through the write port.

The CIM compute macro effectively performs a general matrix multiply (GEMM) operation. However, as input spikes are binary (0 or 1), explicit multiplication is unnecessary and can be replaced by only accumulation operations. Regardless, to perform convolution operations using the hardware capable of performing GEMM efficiently, they are first converted to a GEMM-compatible format, for example using the im2col algorithm to transform the input tensor~\cite{sze2017efficient, chetlur2014cudnn}.

Traditionally, im2col is implemented as a software preprocessing step, requiring additional memory due to data replication. Our design uses the input loader to perform im2col directly in hardware during execution.  The latency of this hardware im2col operation is hidden by exploiting the dual-port SRAM's separate read and write paths. This allows the S2A to begin reading and processing data from the IFspad as soon as the input loader has populated a few rows, rather than waiting for the entire IFspad to be filled. In addition to the  im2col transformation, the input loader handles any necessary zero padding and incorporates stride values directly into the IFspad's data layout. The dataflow mapping utilized by the input loader is detailed in the following section.

\subsection{Reconfigurable operating modes}\label{sec:modes}

\begin{figure}[!t]
\centering
\includegraphics[width=\linewidth]{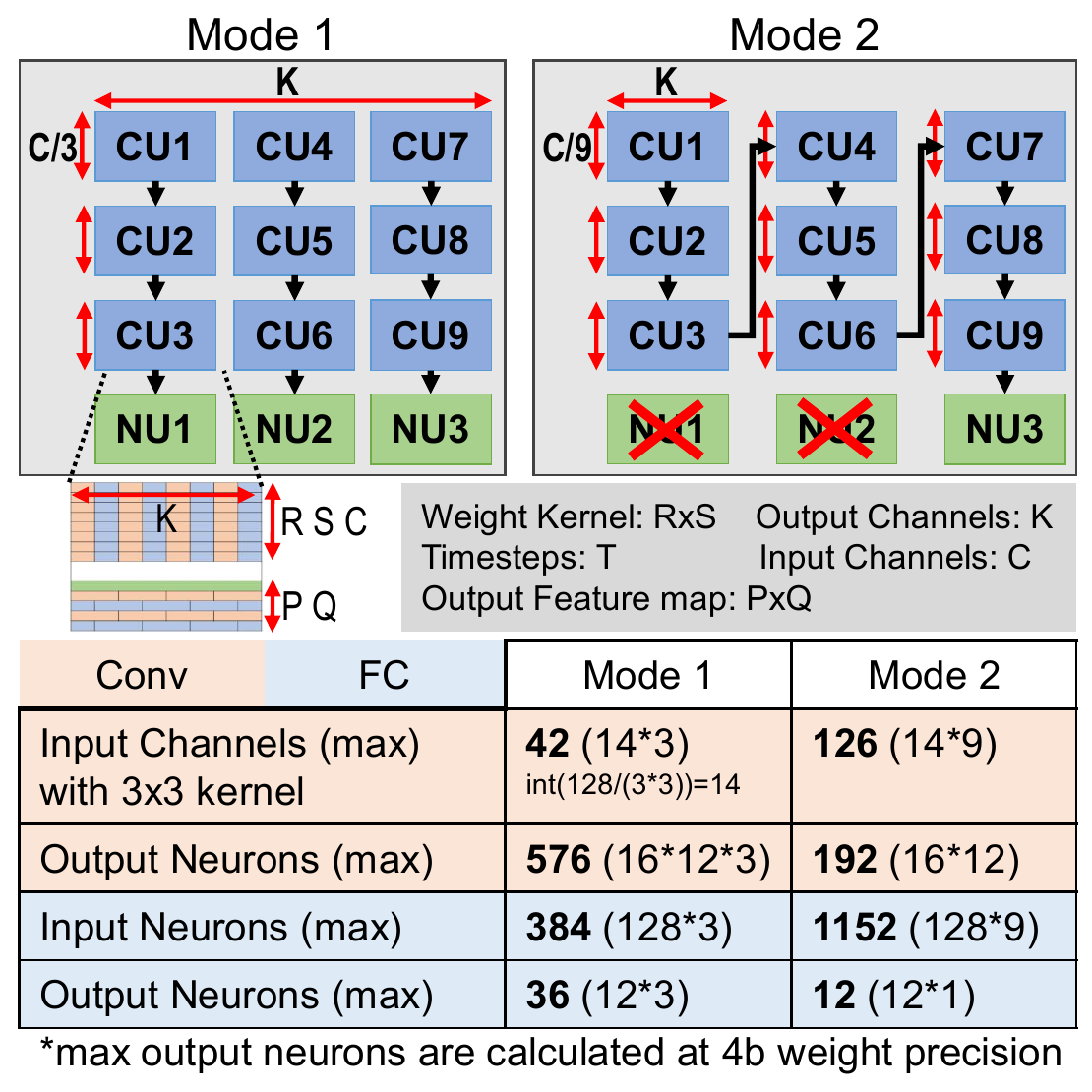}
\caption{Reconfigurable operating modes.}
\label{fig:op_modes}
\end{figure}

% Each macro computes partial Vmems for multiple output neurons concurrently. We use a weight-stationary mapping strategy where K is mapped along the column dimension of the macro, and spatial input dimensions (RxSxC), over which a partial Vmem value is accumulated, are mapped across the rows of a compute macro. We also confine the movement of partial Vmem values within the core by mapping the entire input fan-in (RxSxC for Conv and the number of input neurons for FC layers) directly to the core. This is achieved with the help of two operating modes: mode 1 and mode 2, as shown in Fig.~\ref{fig:op_modes}. Mode 1 suits layers with smaller fan-in, enabling simultaneous output calculations for multiple channels. In contrast, while sacrificing some parallelism, mode 2 minimizes Vmem data movement for layers with larger fan-in. This mapping strategy makes our design easily scalable to a multi-core architecture where each core can process independent output neurons in parallel.

Each compute macro in our accelerator computes partial Vmems for multiple output neurons concurrently. We employ a weight-stationary mapping strategy, common for in-memory computing designs, where different output channels of a Conv layer (\(K\)) or different output neurons for an FC layer are mapped along the column dimension of the macro. The spatial input dimensions (\(R, S, C\)), which represent the receptive field over which a partial Vmem is accumulated, are mapped across the rows of the compute macro. For a 4-bit weight precision, each row can store weights corresponding to \(12\) i.e. \((48/4)\) output neurons. One input spike corresponds to the weight-to-Vmem accumulation of these \(12\) neurons in two cycles (one even and one odd). To utilize the weight reuse in convolution operations, there are \(32\) Vmem rows per macro. Only \(2\) out of these 32 rows are used for  FC layers because there is no weight reuse. For a weight bit precision of \(W_b\), the number of output neurons per macro is represented by,

\begin{equation}\label{x}
    \text{\# output neurons per macro} = \frac{48}{W_b} * 16
\end{equation}

where, \(48\) corresponds to the column dimension of the SRAM array, and \(16\) represents the effective number of Vmem rows \((32/2)\) as there are two rows for a weight row.

Our design limits the movement of partial Vmem values within the core. This is achieved by mapping the entire input fan-in (\(R\times S\times C\) for Conv layers or the number of input neurons for FC layers) onto the core itself. This mapping is facilitated by two operating modes, illustrated in Fig.~\ref{fig:op_modes}:

\begin{itemize}
    \item \textbf{Mode 1:} This mode is used for layers with a smaller input fan-in \((<128\times3)\), where the spatial input dimensions can fit entirely within three compute macros. Mode 1 uses three parallel pipelines each with three compute and one neuron macro to enable simultaneous computation of partial Vmems for more output channels.

    \item \textbf{Mode 2:} In contrast, mode 2 is used for layers with larger input fan-ins \((>128\times3 \text{ and} <128\times9)\). In this mode, partial Vmems are accumulated across \(9\) compute macros, and only \(1\) neuron macro is utilized. This mode sacrifices some parallelism but reduces data movement compared to an alternative where partial Vmems are stored off-chip and need to be fetched for every timestep.
\end{itemize}

The number of output channels processed in parallel by these modes is given by,

\begin{equation}\label{y}
    3 * \frac{48}{W_b} \text{(Mode 1) or } \frac{48}{W_b} \text{(Mode 2)}
\end{equation}

% \begin{figure}[ht]
% \centering
% \includegraphics[width=0.9\linewidth]{figures/SNN_Chip_mapping.pdf}
% \caption{Dataflow mapping for a convolution layer in Mode 1.}
% \label{fig:mapping}
% \end{figure}

% The overall mapping of a convolution layer on our SNN core can be visualized in Fig.~\ref{fig:mapping}.
Our layer mapping strategy, coupled with the adaptable operating modes makes our design easily scalable to a multi-core architecture where each core can process independent output neurons in parallel, increasing throughput without additional data movement. The input loader uses this mapping information to load data into the IFspad for each macro.

\subsection{Timestep pipelining}\label{sec:pipeline}

\begin{figure*}[!t]
\centering
\includegraphics[width=\textwidth]{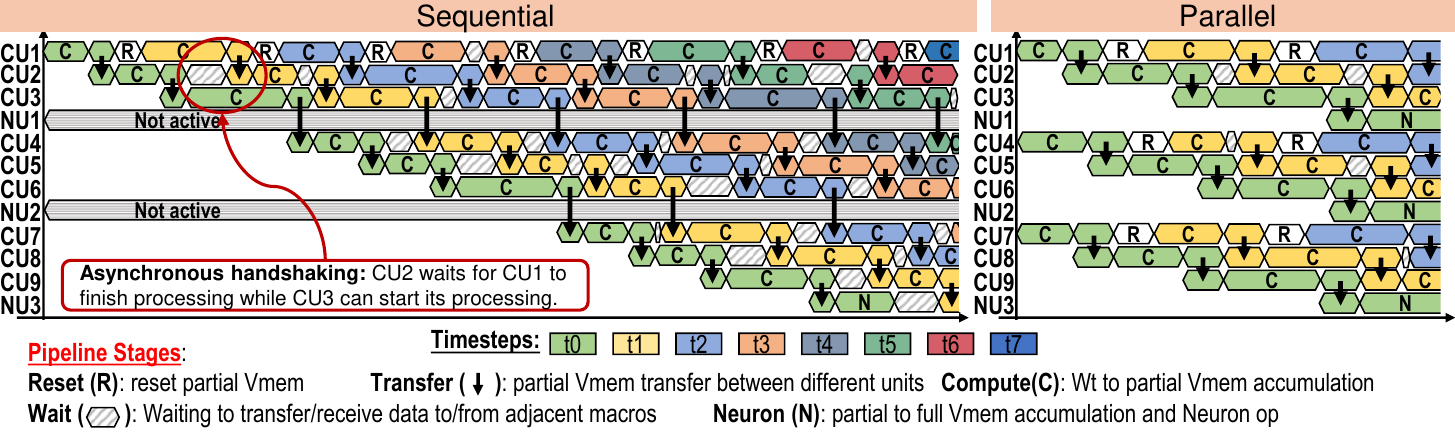}
\caption{Timestep pipelining with asynchronous handshaking.}
\label{fig:pipelining}
\end{figure*}

% We pipelined the timesteps across compute and neuron units (Fig.~\ref{fig:pipelining}), facilitated by an asynchronous handshaking mechanism that coordinates the variable (sparsity-dependent) computation times across different macros. For instance, the compute macro C2 processes partial Vmems for timestep t1 and then passes them to C3, pausing until C1 completes its computations for timestep t2. During this wait, C3 begins the computations for timestep t1. We evenly distribute the input channels among the compute macros, as shown in Fig.~\ref{fig:core_arch}(b) to minimize waiting times.

We pipelined the timesteps across compute and neuron units as demonstrated by Fig.~\ref{fig:pipelining}. Each neuron macro takes a fixed number of cycles which is given by,
\begin{equation}
    \text{\# cycles} = 2*32+2 = 66;
\end{equation}

where \(32*2\) represents the partial-to-full Vmem accumulation and threshold comparison cycles and \(2\) additional cycles are for filling and emptying the column peripheral pipeline.
However, the execution time of each compute macro depends on the sparsity of spikes in its IFspad. Therefore, a pipeline with constant execution times will need to assume the worst-case sparsity and will adversely affect the latency and throughput of the design. Instead, to maintain high computational efficiency despite the variable execution times, we introduce an asynchronous handshaking mechanism to coordinate computations ensuring that the delays are incurred only due to data dependence, and each unit can start computation as soon as it receives the required data.

As illustrated in Fig.~\ref{fig:pipelining}, after the compute unit CU2 processes partial Vmems for timestep t1, it forwards them to CU3 and waits to receive data from CU1. CU3 can start processing its partial Vmems for timestep t1 during this period. Meanwhile, after completing its computations for timestep t2 CU1 will send its data to CU2. This dynamic, asynchronous coordination ensures efficient utilization of resources and minimizes the computational pipeline stalls.

To further optimize this pipelined approach, the input channels are evenly distributed among the compute macros, as shown in Fig.~\ref{fig:op_modes}, utilizing the same number of rows in each macro. This balanced distribution minimizes the potential waiting times, as the only factor responsible for the variable execution time of each macro is the variation in spike density and not the number of valid data rows.

\begin{figure}[!t]
\centering
\includegraphics[width=0.7\linewidth]{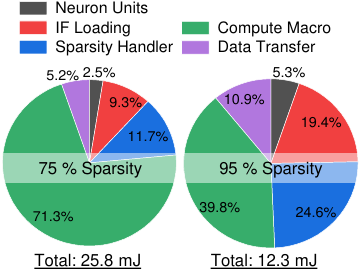}
\caption{Peripheral switching Overhead.}
\label{fig:energy_dist}
\end{figure}

\section{Experiments and Results}\label{sec:experiments}
% Fig~\ref{fig:energy_dist} characterizes the energy distribution for each component within the proposed SNN core, at input sparsities of 75\% and 95\%. Notably, CIM macros dominate the energy consumption even at very high input sparsity (95\%). This suggests that the energy overhead associated with control logic does not overpower the energy required for actual computations (Compute Macro and Neuron Units). Moreover, the energy consumed by data movement constitutes only a small portion of the total energy consumption in our design. Note that the total energy consumption decreases by more than 50\% when the input sparsity increases from 75\% to 95\%.

Fig.~\ref{fig:energy_dist} presents a breakdown of the energy consumption for each component within the SNN core at two input sparsity levels: 75\% and 95\%. Notably, the CIM macros (Compute Macro + Neuron Units), responsible for weight-to-Vmem accumulation, and neuron operations are the dominant energy consumers both at a moderate input sparsity of 75\% and a very high input sparsity of 95\%. This observation highlights the efficiency of our design, as the energy overhead associated with control logic and peripheral circuitry does not overpower the energy required for actual computations. However, the overall energy consumption of the SNN core is significantly influenced by input sparsity, decreasing by more than 50\% when the input sparsity increases from 75\% to 95\%, highlighting the importance of exploiting input sparsity in SNNs.

Furthermore, the energy consumed by data movement, a critical bottleneck in many traditional architectures, constitutes only a small fraction of the total energy in our design. This can be attributed to processing in-memory and our design's optimized dataflow, which restricts data movement. Moreover, these trends will be consistent for different layer sizes as a result of our reconfigurable operating modes.

\begin{figure}[!t]
\centering
\includegraphics[width=0.55\linewidth]{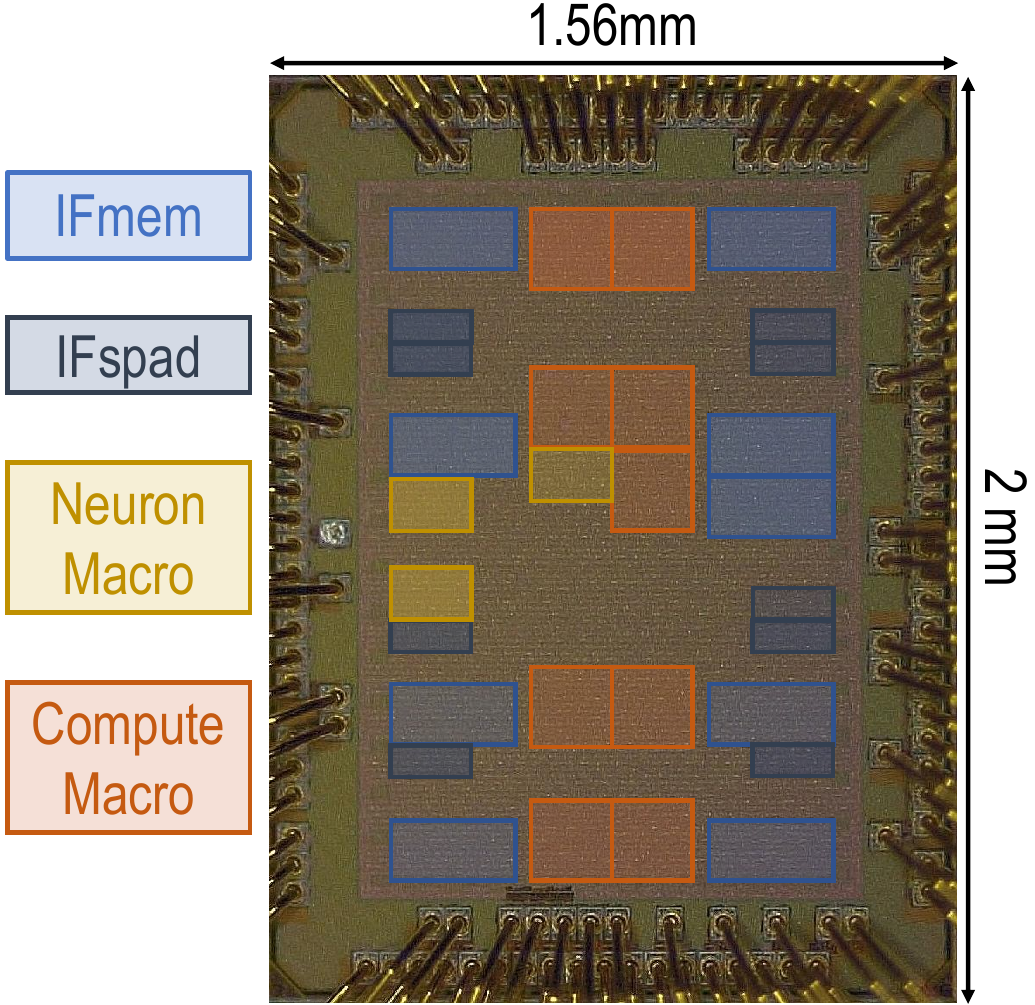}
\caption{Die Micrograph.}
\label{fig:chip}
\end{figure}

{
\setlength{\tabcolsep}{2pt} % Reduce horizontal padding
\begin{table}
    \centering
    \caption{Chip Summary}
    \label{tab:chip_summary}
    \resizebox{\linewidth}{!}{
    \begin{tabular}{|>{\centering\arraybackslash}m{3.6cm}|>{\centering\arraybackslash}m{0.7cm}|>{\centering\arraybackslash}m{0.7cm}|>{\centering\arraybackslash}m{0.7cm}|>{\centering\arraybackslash}m{0.7cm}|>{\centering\arraybackslash}m{0.7cm}|>{\centering\arraybackslash}m{0.7cm}|} \hline
         \textbf{Technology}                 & \multicolumn{6}{{|c|}}{\qty{65}\nm~CMOS} \\ \hline
         \textbf{Die size}                   & \multicolumn{6}{{|c|}}{1.56mm x 2mm}\\ \hline
                                             & \multicolumn{6}{|c|}{Total: 52.08 kB} \\ 
         \textbf{On chip SRAM}               & \multicolumn{6}{|c|}{IMC Macros: 9.7 kB} \\
                                             & \multicolumn{6}{|c|}{Excluding input spike memories*: 12.7 kB}\\ \hline
         \textbf{Supply Voltage}             & \multicolumn{6}{|c|}{0.9V - 1.2V} \\ \hline
         \textbf{Frequency}                  & \multicolumn{6}{|c|}{50MHz - 150MHz} \\ \hline
         \textbf{Weight/Vmem Precision}      & \multicolumn{6}{|c|}{4/7-bit, 6/11-bit, 8/15-bit} \\ \hline
                                             & \multicolumn{3}{|c|}{@50MHz, 0.9V} & \multicolumn{3}{|c|}{@150MHz, 1V}\\ \hline
         \textbf{Power Consumption (mW)}     & \multicolumn{3}{|c|}{4.9}    & \multicolumn{3}{|c|}{18}\\ \hline
         \textbf{Weight Precision}           & {\textbf{4b}} & {\textbf{6b}} & {\textbf{8b}} & {\textbf{4b}} & {\textbf{6b}} & {\textbf{8b}} \\ \hline
         \textbf{Energy Efficiency (TOPS/W)} &&&&&& \\
         \textbf{@95\% sparsity}             & {5}           & {3.34}        & {2.5}         & {4.09}        & {2.73}        & {2.04} \\ \hline
         \textbf{Throughput (GOPS)}          &&&&&& \\
         \textbf{@95\% sparsity}             & {24.54}       & {16.36}       & {12.27}       & {73.59}       & {49.06}       & {36.80} \\ \hline

         \multicolumn{7}{l}{*Large input spike memories are not required if the core is used} \\
         \multicolumn{7}{l}{as part of a larger system and data is streamed in continuously.} \\
    \end{tabular}
    }
\end{table}
}

Fig.~\ref{fig:chip} shows the die micrograph of the proposed SNN core and Table \ref{tab:chip_summary} provides a comprehensive summary of its specifications. Fabricated using \qty{65}\nm~CMOS technology, the chip occupies an area of \qty{3.12}{\mm\squared} including the I/O pads. It features a total on-chip SRAM of \qty{52.08}{\kilo\byte}, with \qty{9.7}{\kilo\byte} allocated to IMC macros and \qty{39.38}{\kilo\byte} allocated to the input memories (IFmem). We kept our IFmems large to test the functionality and fit the inputs corresponding to large layers on-chip. In a more practical system, where the IFmems can be written in parallel to the operations, or when this SNN core is used as a part of a larger system, the IFmem size can be reduced significantly as we do not need to store all the inputs at the same time on-chip. The chip operates within a voltage range of \qty{0.9}{\volt} to \qty{1.2}{\volt} and a frequency range of \qty{50}{\mega\hertz} to \qty{150}{\mega\hertz}. It consumes \qty{4.9}{\milli\watt} power at \qty{50}{\mega\hertz} and \qty{0.9}{\volt}, and \qty{18}{\milli\watt} power at \qty{150}{\mega\hertz} and \qty{1}{\volt}. The chip's energy efficiency, measured in Tera operations per second per watt (\unit{TOPS/W}), varies with weight precision and sparsity levels, reaching up to 5 \unit{TOPS/W} at 4-bit weight precision and 95\% sparsity. Its throughput, measured in Giga operations per second (\unit{GOPS}), peaks at 73.59 \unit{GOPS} at 4-bit weight precision, 150MHz, and 95\% sparsity.

{
\setlength{\tabcolsep}{2pt} % Reduce horizontal padding
\begin{table}[tbp]
    \centering
    \caption{Network Details}
    \label{tab:network_details}
    \resizebox{\linewidth}{!} {
    \begin{tabular}{|>{\raggedright\arraybackslash}m{1.5cm}||>{\centering\arraybackslash}m{1cm}|>{\centering\arraybackslash}m{1.2cm}|>{\centering\arraybackslash}m{0.9cm}|>{\centering\arraybackslash}m{1.5cm}|>{\centering\arraybackslash}m{0.9cm}|} \hline
         Application                        & Input size & Timesteps & Input layer & Intermediate layers & Output layer \\ \hline \hline
         Optical flow estimation            & 288x384    & 10        & Conv (2,32) & 6*Conv (32,32)      & Conv (32,2) \\ \hline 
         Gesture recognition\textcolor{orange}{$^{\mathrm{a}}$} & 64x64      & 20        & Conv (2,16) & 4*Conv (16,16)      & FC (64,11) \\ \hline

         \multicolumn{6}{l}{Conv (Input,Output Channels), FC (Input,Output Neurons) } \\
         \multicolumn{6}{l}{\textcolor{orange}{$^{\mathrm{a}}$}2x2 maxpool with stride 2 after every two intermed. conv layers.}
    \end{tabular}
    }
\end{table}
}

\begin{figure}[!t]
\centering
\includegraphics[width=\linewidth]{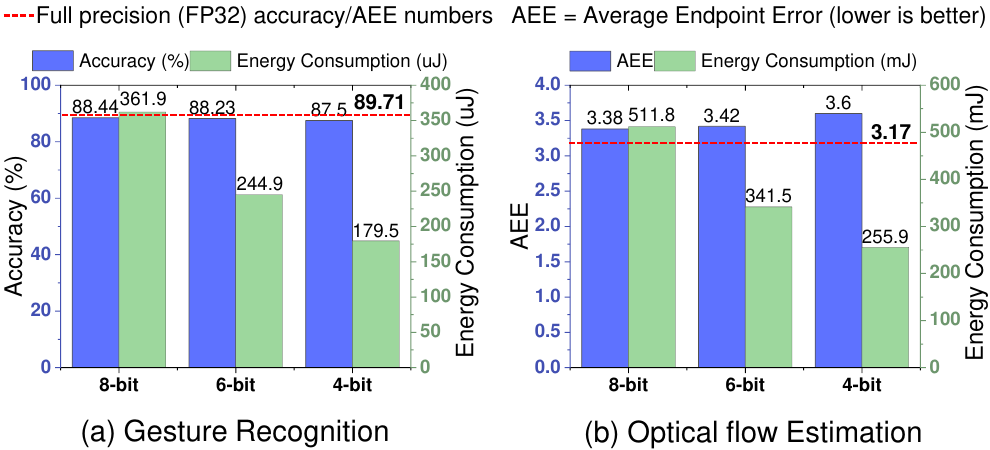}
\caption{Accuracy and energy consumption trade-off at different weight precisions for gesture recognition and optical flow estimation.}
\label{fig:acc_vs_efficiency}
\end{figure}

\begin{figure}[!t]
\centering
\includegraphics[width=\linewidth]{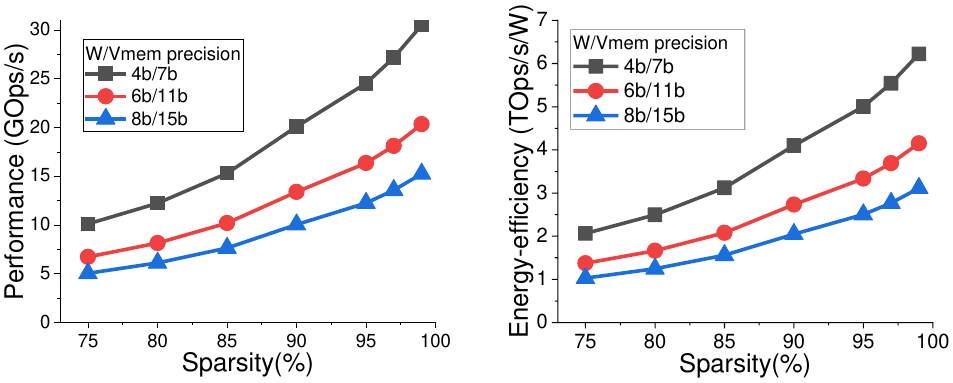}
\caption{Peak performance and energy efficiency trend with changing sparsity and bit precision}
\label{fig:perf_energy_eff}
\end{figure}

% Fig.~\ref{fig:perf_energy_eff} and Fig.~\ref{fig:acc_vs_efficiency} show the measured results of our chip. We evaluated the SNN core on two applications: hand gesture recognition on the IBM DVS gestures dataset~\cite{amir2017low} and optical flow estimation on the DSEC flow dataset~\cite{gehrig2021dsec}. Table~\ref{tab:network_details} shows the network details, and Fig.~\ref{fig:acc_vs_efficiency} shows the energy and accuracy trade-off with changing bit precision for these tasks at 50MHz and 0.9V. Since this is a digital CIM design, there is no loss in accuracy at hardware implementation. Average endpoint error (AEE) is commonly used as a performance metric for optical flow estimation~\cite{negi2023best} as opposed to accuracy in gesture recognition. Fig.~\ref{fig:perf_energy_eff} shows the trend for peak performance (GOPS) and energy efficiency (TOPS/W) with changing sparsity and bit precision. For instance, there is a 2x improvement in throughput when the precision changes from 8-bit to 4-bit at the same sparsity or when the sparsity changes from 80\% to 95\% at 4-bit precision.

% Please add the following required packages to your document preamble:
% \usepackage{multirow}
{
\setlength{\tabcolsep}{2pt} % Reduce horizontal padding
\begin{table*}[h]
    \centering
    \caption{Comparison Table}
    \label{tab:comparison}
    \resizebox{0.96\textwidth}{!}{
    \begin{tabular}{|c|c|c|c|c|c|c|} \hline
        \textbf{Parameter}                                                                                         & \begin{tabular}[c]{@{}c@{}}\textbf{\chipname}\\ \textbf{(This Work)}\end{tabular}                                                                & \begin{tabular}[c]{@{}c@{}}\textbf{C-DNN}\\ \textbf{ISSCC’23\cite{kim2023c}}\end{tabular}                                 & \begin{tabular}[c]{@{}c@{}}\textbf{ANP-I}\\ \textbf{ISSCC’23\cite{zhang202322}}\end{tabular}                                                             & \begin{tabular}[c]{@{}c@{}}\textbf{Reck-On}\\ \textbf{ISSCC’22\cite{frenkel2022reckon}}\end{tabular}                                                                                           & \begin{tabular}[c]{@{}c@{}}\textbf{\(\mu\)Brain}\\ \textbf{Frontiers’21\cite{stuijt2021mubrain}}\end{tabular}                                     & \textbf{\begin{tabular}[c]{@{}c@{}}\textbf{SD Training}\\ \textbf{ISSCC’19\cite{park20197}}\end{tabular}} \\ \hline
        \begin{tabular}[c]{@{}c@{}}\textbf{Technology}\\ \textbf{Supply (V)}\\ \textbf{Freq. (MHz)}\\ \textbf{Area\textcolor{orange}{$^{\mathrm{+}}$} (mm2)}\end{tabular}       & \begin{tabular}[c]{@{}c@{}}\qty{65}\nm~CMOS\\ 0.9 - 1.2\\ 50 - 150\\ 3.12\end{tabular}                                    & \begin{tabular}[c]{@{}c@{}}28nm CMOS\\ 0.7 – 1.1\\ 50 - 200\\ 20.25\end{tabular}                   & \begin{tabular}[c]{@{}c@{}}28nm CMOS\\ 0.56 – 0.9\\ 40 – 210\\ 1.63\end{tabular}                                               & \begin{tabular}[c]{@{}c@{}}28nm FDSOI\\ 0.5 – 0.8\\ 13-115\\ 0.87\end{tabular}                                                                                 & \begin{tabular}[c]{@{}c@{}}40nm CMOS\\ 1.1\\ -\\ 2.82\end{tabular}                                          & \begin{tabular}[c]{@{}c@{}}\qty{65}\nm~CMOS\\ 0.8\\ 20\\ 10.08(core)\end{tabular}        \\ \hline
       \textbf{ Target Domain}                                                                                     & SNN                                                                                                                & ANN/SNN                                                                                            & SNN                                                                                                                            & Spiking RNN                                                                                                                                                    & SNN                                                                                                         & SNN                                                                               \\ \hline
        \textbf{Neuron Model}                                                                                      & \textcolor{blue}{Flexible}                                                                                                           & Fixed                                                                                              & Fixed                                                                                                                          & Fixed                                                                                                                                                          & Flexible\textcolor{orange}{$^{\mathrm{1}}$}                                                                                                   & Fixed                                                                             \\ \hline
        \textbf{Neurons}                                                                                           & \begin{tabular}[c]{@{}c@{}}\textcolor{blue}{inp 1152\textcolor{orange}{$^{\mathrm{a}}$}}\\ \textcolor{blue}{op 576\textcolor{orange}{$^{\mathrm{b}}$}}\end{tabular}                                                        & op: 2048\textcolor{orange}{$^{\mathrm{c}}$}                                                                                          & (1024)-512-10                                                                                                                  & (256)-256-16                                                                                                                                                   & (256)-64-16                                                                                                 & 2x200(hidden)                                                                     \\ \hline
        \textbf{Compute type}                                                                                      & Digital CIM                                                                                                        & Digital                                                                                            & Async. Digital                                                                                                                 & Async Digital                                                                                                                                                  & Async Digital                                                                                               & Digital                                                                           \\ \hline
        \begin{tabular}[c]{@{}c@{}}\textbf{Weight Prec.}\\ \textbf{Vmem Prec.}\end{tabular}                                 & \begin{tabular}[c]{@{}c@{}}\textcolor{blue}{4 / 6 / 8} \\ \textcolor{blue}{7 / 11 / 15}\end{tabular}                                                   & \begin{tabular}[c]{@{}c@{}}4/8\\ -\end{tabular}                                                    & \begin{tabular}[c]{@{}c@{}}hidden: 8, op: 10\\ -\end{tabular}                                                                  & \begin{tabular}[c]{@{}c@{}}8\\ 16\end{tabular}                                                                                                                 & \begin{tabular}[c]{@{}c@{}}4\\ 7\end{tabular}                                                               & \begin{tabular}[c]{@{}c@{}}-\\ 8\end{tabular}                                     \\ \hline
        \begin{tabular}[c]{@{}c@{}}\textbf{Energy Eff.} \\ {\textbf{[}}\textbf{TOPS/W}{\textbf{]}} \\  \textbf{(default)}\end{tabular} & \begin{tabular}[c]{@{}c@{}}4b wt: 5 (26.95)\textcolor{orange}{$^{\mathrm{d}}$}\\ 6b wt: 3.34 (18)\textcolor{orange}{$^{\mathrm{d}}$}\\ 8b wt: 2.5 (13.5)\textcolor{orange}{$^{\mathrm{d}}$}\\ @50MHz, 0.9V\end{tabular} & \begin{tabular}[c]{@{}c@{}}CIFAR10: 63.3\textcolor{orange}{$^{\mathrm{e}}$}\\ ImageNet: 20.8\textcolor{orange}{$^{\mathrm{e}}$}\\ @50MHz, 0.7V\end{tabular}           & \begin{tabular}[c]{@{}c@{}}1.5 pJ/SOP\\ @40MHz, 0.56V\end{tabular}                                                          & \begin{tabular}[c]{@{}c@{}}5.3 pJ/SOP\\ @13MHz, 0.5V\end{tabular}                                                                                             & \begin{tabular}[c]{@{}c@{}}MNIST\\ 308nJ/prediction\\ (160nJ/prediction)\textcolor{orange}{$^{\mathrm{d}}$}\\ @1.1V\end{tabular}              & \begin{tabular}[c]{@{}c@{}}3.42\\ (18.43)\textcolor{orange}{$^{\mathrm{d}}$}\\ @20MHz 0.8V\end{tabular}           \\ \hline
        \begin{tabular}[c]{@{}c@{}}\textbf{Tasks and}\\ \textbf{Datasets}\end{tabular}                                      & \begin{tabular}[c]{@{}c@{}}IBM DVS \\ gest. classif.\\ \textcolor{blue}{Optical flow}\\ \textcolor{blue}{estimation on}\\ \textcolor{blue}{DSEC-flow}\end{tabular}       & \begin{tabular}[c]{@{}c@{}}Image classif. on\\ CIFAR-10/100\\ and ImageNet\end{tabular}            & \begin{tabular}[c]{@{}c@{}}MNIST classif.\\ IBM DVS\\ gesture classif.\\ Keyword\\ spotting (N-\\ TIDIGIT, SeNic)\end{tabular} & \begin{tabular}[c]{@{}c@{}}IBM DVS\\ gesture classif.\\ Keyword spotting\\ (Spiking\\ Heidelberg Digits)\\ Navigation (delayed\\ cue integration)\end{tabular} & \begin{tabular}[c]{@{}c@{}}MNIST classif.\\ Radar hand gest.\\ classif. (customized\\ dataset)\end{tabular} & \begin{tabular}[c]{@{}c@{}}MNIST\\ classif.\end{tabular}                          \\ \hline
        \begin{tabular}[c]{@{}c@{}}\textbf{Reconfigurable}\\ \textbf{Network Arch.}\end{tabular}                                    & \textcolor{blue}{Yes\textcolor{orange}{$^{\mathrm{f}}$}}                                                                                                               & Yes\textcolor{orange}{$^{\mathrm{g}}$}                                                                                               & No                                                                                                                             & No                                                                                                                                                             & No                                                                                                          & No                                                                                \\ \hline
        \begin{tabular}[c]{@{}c@{}}\textbf{Modified} \\ \textbf{Training}\end{tabular}                                      & \textcolor{blue}{No}                                                                                                                 & Yes                                                                                                & Yes                                                                                                                            & Yes                                                                                                                                                            & No                                                                                                          & Yes                                                                               \\ \hline
        \begin{tabular}[c]{@{}c@{}}\textbf{Sparsity} \\ \textbf{Support}\end{tabular}                                       & \begin{tabular}[c]{@{}c@{}}\textcolor{blue}{Supports} \\ \textcolor{blue}{unstructured} \\ \textcolor{blue}{input sparsity}\end{tabular}                                 & \begin{tabular}[c]{@{}c@{}}\textcolor{orange}{$^{\mathrm{g}}$}Spiking core \\ used when \\ sparsity\textgreater{}97.7\%\end{tabular} & \begin{tabular}[c]{@{}c@{}}Yes\\ (event-driven)\end{tabular}                                                                   & \begin{tabular}[c]{@{}c@{}}Yes\\ (event-driven)\end{tabular}                                                                                                   & \begin{tabular}[c]{@{}c@{}}Yes\\ (event-driven)\end{tabular}                                                & \begin{tabular}[c]{@{}c@{}}Yes (propagate\\ only spikes)\end{tabular}             \\ \hline
        
        \multicolumn{7}{l}{\textcolor{orange}{$^{\mathrm{+}}$}die area as majority of the other works report only die area,  \textcolor{orange}{$^{\mathrm{a}}$}max input neurons when processing FC in mode 2,} \\
        \multicolumn{7}{l}{\textcolor{orange}{$^{\mathrm{b}}$}max output neurons when processing conv in mode 1, \textcolor{orange}{$^{\mathrm{c}}$}4 clusters*8 snn cores*64 neurons per core, } \\
         \multicolumn{7}{l}{\textcolor{orange}{$^{\mathrm{d}}$}scaled to 28nm for comparison with other accelerators assuming \(energy \propto tech^2\), \textcolor{orange}{$^{\mathrm{e}}$}only SNN processing numbers,} \\
        \multicolumn{7}{l}{\textcolor{orange}{$^{\mathrm{f}}$}has two operating modes to trade-off fan-in with fan-out depending on application requirement,} \\
         \multicolumn{7}{l}{ \textcolor{orange}{$^{\mathrm{g}}$}each layer’s mapping is decided based on sparsity, \textcolor{orange}{$^{\mathrm{1}}$}supports IF but LIF can be used with an external clock} \\

        \end{tabular}
}
\end{table*}

}

We evaluate the SNN core on two applications: hand gesture recognition on the IBM DVS gestures dataset~\cite{amir2017low} and optical flow estimation on the DSEC flow dataset~\cite{gehrig2021dsec}. Table~\ref{tab:network_details} provides detailed information on the network architectures employed for each task. Fig.~\ref{fig:acc_vs_efficiency} and Fig.~\ref{fig:perf_energy_eff} present the measured results of our chip. The energy-accuracy trade-off at varying bit precision is shown in Fig.~\ref{fig:acc_vs_efficiency} at \qty{50}{\mega\hertz} and \qty{0.9}{\volt}. It is crucial to note that since this is a digital CIM design, there is no loss in accuracy at hardware implementation. Average endpoint error (AEE) is commonly used as a performance metric for optical flow estimation~\cite{negi2023best} as opposed to accuracy.

Fig.~\ref{fig:perf_energy_eff} shows the trend for peak performance (\unit{GOPS}) and energy efficiency (\unit{TOPS/W}) as a function of input sparsity and weight precision. For instance, there is a \(2\times\) improvement in throughput when the precision changes from 8-bit to 4-bit at the same sparsity or when the sparsity changes from \(80\%\) to \(95\%\) for 4-bit weight precision.

\section{Discussion}\label{sec:discussion}
Table~\ref{tab:comparison} compares \chipname with contemporary digital SNN accelerators. Although asynchronous designs with fixed network architectures~\cite{zhang202322, frenkel2022reckon, stuijt2021mubrain} consume very low power for specific applications, their lack of scalability and reconfigurability makes them unsuitable for a wide range of other applications. In contrast, \chipname offers advanced dataflow and reconfigurability, catering to a wide range of SNN applications. Additionally, hybrid designs are mainly suitable for applications with analog inputs where ANNs outperform SNNs in accuracy~\cite{kim2023c, liu2023aa}, while for dynamic tasks such as optical flow estimation, SNNs outperform ANNs~\cite{negi2023best} and our design targets these applications. Moreover, our design does not require any modified training methodology and can adapt to a variety of workload specifications.

A parallel research direction to this work is developing analog CIM architectures \cite{kim2022neuro, liu2023aa} for high throughput and energy efficiency using CMOS and post-CMOS technologies such as memristors. However, analog CIM needs efficient mechanisms to address various non-idealities. Moreover, using ANNs and SNNs to complement each other for various robotic applications is another compelling area of research \cite{chang202373, vidal2018ultimate}. Authors in \cite{chang202373} not only combine ANNs and SNNs but also use both RRAM CIM and SRAM CNM for real-time target identification and tracking. 
Future research directions include developing heterogeneous systems that integrate multiple specialized yet reconfigurable architectures to handle diverse computational tasks in real time, adapting to ever-changing algorithmic demands. Concurrently, creating more sophisticated algorithms to leverage the full potential of such reconfigurable heterogeneous hardware platforms can pave the way for highly efficient designs.

\section{Conclusion}\label{sec:conclusion}
This paper presents a scalable and reconfigurable digital CIM SNN accelerator \chipname for dynamic vision tasks such as optical-flow estimation. \chipname introduces a set of key features, including in-memory computations, reconfigurable operating modes, multi-bit precision support, a zero-skipping mechanism for sparse inputs, and an asynchronous handshaking mechanism. These features enable \chipname to adapt to different workloads with varying neuron models, bit precisions, and network architectures while minimizing data movement and energy consumption. \chipname was fabricated in TSMC \qty{65}\nm~CMOS LP technology and demonstrates performance competitive to other digital SNN accelerators proposed in the recent literature even with advanced reconfigurability. It achieves an energy efficiency of up to 5 TOPS/W at an input sparsity of 95\%, weight precision of 4 bits, and Vmem precision of 7 bits.

\bibliographystyle{IEEEtran}
\bibliography{IEEEabrv,references}

\vfill

\end{document}